\documentclass{ws-ijmpe}

\usepackage[super,compress]{cite}

\usepackage{bm}
\usepackage{graphicx}
\usepackage{amssymb}
\usepackage{epsfig}%
\usepackage{hyperref}
\usepackage{mathrsfs}
\usepackage{multirow}
\usepackage{footmisc}
\usepackage{threeparttable}
\usepackage{booktabs}
\usepackage{appendix}
\usepackage{float}
\usepackage{color}

\begin{document}

\markboth{Y. Cui et al.}{The influence of $\delta$ meson on the isospin splitting of in-medium $NN\to N\Delta$ cross sections}

\catchline{}{}{}{}{}

\title{The influence of $\delta$ meson on the isospin splitting of in-medium $NN\to N\Delta$ cross sections}

\author{Ying Cui}

\address{China Institute of Atomic Energy\\
 Beijing 102413, China\\
yingcuid@163.com}

\author{Yingxun Zhang}
\address{China Institute of Atomic Energy\\
 Beijing 102413, China\\
zhyx@ciae.ac.cn}
\address{Guangxi Key Laboratory Breeding Base of Nuclear Physics and Technology, Guangxi Normal University\\
 Guilin 541004, China}

\author{Yuan Tian}
\address{China Institute of Atomic Energy\\
 Beijing 102413, China}

\author{Zhuxia Li}
\address{China Institute of Atomic Energy\\
 Beijing 102413, China}

\maketitle

\begin{history}
\received{Day Month Year}
\revised{Day Month Year}
\end{history}

\begin{abstract}
The isospin splitting of the in-medium $NN\rightarrow N\Delta$ cross sections in  asymmetric nuclear medium are investigated in the framework of the one-boson exchange model by including  $\delta$ and $\rho$ mesons. Our results show that  the medium correction factors $R=\sigma_{ NN\rightarrow N\Delta}^*/\sigma_{NN\rightarrow N\Delta}^{\text{free}}$ have $R_{pp \to n\Delta ^{++}} < R_{nn \to p\Delta ^{-}}$ and $R_{NN \to N\Delta ^{+}} <R_{NN \to N\Delta ^{0}}$ by using the without-$\delta$ sets.  By including the $\delta$ meson, it appears the  totally opposite results in the $R$ for different channels, i.e., $R_{pp \to n\Delta ^{++}} > R_{nn \to p\Delta ^{-}}$ and $R_{NN \to N\Delta ^{+}} >R_{NN \to N\Delta ^{0}}$.
\end{abstract}


\ccode{PACS numbers: 24.10. Cn, 25.70.-z}


\section{Introduction}
\label{into}
Studying the medium correction of $NN$ cross sections is an important issue in the field of intermediate energy  heavy ion collisions (HIC), since it can influence the predictions of reaction dynamics, collective flow, stopping power, and particle productions\cite{Chen1998, JYLiu01, QingfengLi2001, zhang07,lehaut10,lopez14,lwchen00,YJWang16} . At the beam energy above about 270 MeV/nucleon, the $\Delta$ resonances appear due to $NN\to N\Delta$ process. Most of the $\Delta$ are produced at the early stage of heavy ion collisions, and then decay into the nucleons and pions.  Thus, the pion is thought as a probe which carries the information of equation of state(EOS) at suprasaturation density. Especially, the pion production and the ratio of the yields of the charged pion $\pi^-/\pi^+$ are proposed to probe the symmetry energy at high density region.

In the transport model simulations, the key ingredient for the $\pi-N-\Delta$ loops is the in-medium $NN\to N\Delta$ cross sections.
The in-medium  $NN\rightarrow N\Delta$  cross sections in symmetric nuclear matter have been investigated by using microscopic approaches \cite{Haar1987c, Haar1987r, Mao1994c, Mao1994l, Larionov2001, Larionov2003, QingfengLi2000} , where the cross sections of different channels  are
obtained by  using of Clebsch-Gordon coefficients for , i.e., $\sigma^*_{pp\rightarrow n\Delta^{++}}:\sigma^*_{nn\rightarrow p\Delta^{-}}: \sigma^*_{pp\rightarrow p\Delta^{+}} :\sigma^*_{pn\rightarrow n\Delta^{+}}: \sigma^*_{pn\rightarrow p\Delta^{0}}:\sigma^*_{nn\rightarrow n\Delta^{0}}=3:3:1:1:1:1$. The medium correction factor $R=\sigma^*_{NN\rightarrow N\Delta}/\sigma^{\text{free}}_{NN\rightarrow N\Delta}$ are exactly the same for each channel.
For the isospin asymmetric nuclear matter, the isospin dependence of $\sigma^*_{NN\to N\Delta}$ and $R$ becomes important.
Li and Li studied the elementary two-body $NN\to N\Delta$ cross section without considering the mass distribution of $\Delta$ resonance and threshold effects by using the same interaction Lagrangian that was used in deriving the collision term of the relativistic
Boltzmann-Uehling-Uhlenbeck (RBUU) microscopic transport theory based on the closed time-path Green's function technique in Ref.~\cite{QingfengLi2017}. Their results show that the  $R$  depends on the isospin channels of $NN\rightarrow N\Delta$ due to the effective masses splitting of N and $\Delta$, which are from the isospin splitting of the scalar self-energies of N (n and p) and $\Delta$s ($\Delta^{++}, \Delta^{+}, \Delta^{0} $ and $\Delta^{-}$). The threshold effect (i.e., energy conservation) and the mass distribution of $\Delta$ resonance on the $\sigma^*_{NN\rightarrow N\Delta}$ in asymmetric nuclear matter has been investigated in Ref.~\cite{Cui2018}. And those effects cause the isospin splitting of $R$ tends to vanish at high energy.

Another important issue is how the $R$ depends on the relativistic mean field (RMF) parameters. In our recent paper \cite{Cui2019} , we analyzed the relation between the in-medium $NN\to N\Delta$ cross section and the RMF parameter sets, especially on the slope of the symmetry energy $L$. Our analyses show that $R$ increase with $L$, but the analyses are limited with $\delta$ meson sets at certain isospin asymmetry in Ref.~\cite{QingfengLi2017,Cui2019} , where the scalar self-energies (from $\delta$ meson) and vector self-energies (from $\rho$ meson) both make the contributions to the isospin splitting of $R_{NN\to N\Delta}$. While, 95\% of RMF parameter sets do not include the $\delta$ meson, and the transport models, such as the Boltzmann-Uehling-Uhlenbeck (BUU) \cite{Larionov2003}/relativistic Vlasov-Uehling-Uhlenbeck (RVUU) \cite{Song2015} have used the RMF sets without/with $\delta$ meson to study the influence of different stiffness of the symmetry energy on the pion observables. Thus, it is worth understanding the influence of $\delta$ meson parameter sets on the isospin splitting of in-medium $NN\to N\Delta$ cross sections by comparing the results from the with  and without  $\delta$ meson parameter sets.

The paper is organized as follows.
The  model for calculation of the in-medium $NN\rightarrow N\Delta$ cross section are briefly described in Sec.~\ref{model}.
In Sec.~\ref{xs}, we discuss the splitting of isospin dependent  medium correction factor $R$ of $NN\rightarrow N\Delta$ cross sections from the sets without and with $\delta$ meson.  And a summary is given in Sec.~\ref{summary}.

\section{The Model}
\label{model}

For the calculation of the in-medium $NN\rightarrow N\Delta$ cross section in isospin asymmetric nuclear matter, we use the one-boson exchange model with the relativistic Lagrangian including nucleon and $\Delta$ ($\Delta$ is the  Rarita-Schwinger spinor of spin-3/2 \cite{Huber1994,Machleidt1987,Benmerrouche1989}).

\subsection{Effective Lagrangian}
In this paper, two types of RMF parameter sets are adopted for estimating the in-medium cross section: one type is the without-$\delta$ meson sets, i.e., the NL$\rho$-$\Delta$\cite{Liu2005}
 , DDME2-$\Delta$\cite{Lalazissis2005} and  DDRH$\rho$-$\Delta$ \cite{Gaitanos2004} , and another type is the with-$\delta$ meson sets, i.e.,  the  NL$\rho\delta$-$\Delta$\cite{Liu2005} , DDME$\delta$-$\Delta$ \cite{Roca2011} and  DDRH$\rho\delta$-$\Delta$\cite{Gaitanos2004} .

Including the $\sigma$, $\omega$, $\rho$, $\delta$, the unified  Lagrangian  reads:
\begin{equation}
\label{Lag}	
\mathcal{L}=\mathcal{L}_I+\mathcal{L}_F,
\end{equation}
where $\mathcal{L}_F$ is
\begin{eqnarray}
\label{lag_f}
\mathcal{L}_{F}=&&\bar{\Psi}[i\gamma_{\mu}\partial^{\mu}-m_{N}]\Psi+\bar{\Delta}_{\lambda}[i\gamma_{\mu}\partial^{\mu}-m_{\Delta}]\Delta^{\lambda}\\
&&+\frac{1}{2}\left(\partial_{\mu}\sigma\partial^{\mu}\sigma-m_{\sigma}^2\sigma^2\right)-U(\sigma)\nonumber\\
&&-\frac{1}{4}\omega_{\mu\nu}\omega^{\mu\nu}+\frac{1}{2}m^{2}_{\omega}\omega_{\mu}\omega^{\mu}\nonumber\\
&&+\frac{1}{2}\left(\partial_{\mu}\bm{\pi}\partial^{\mu}\bm{\pi}-m^{2}_{\pi}\bm{\pi}^{2}\right)-\frac{1}{4}\bm{\rho}_{\mu\nu}\bm{\rho}^{\mu\nu}+\frac{1}{2}m^{2}_{\rho}\bm{\rho}_{\mu}\bm{\rho}^{\mu}\nonumber\\
&&+\frac{1}{2}\left(\partial_{\mu}\bm{\delta}\partial^{\mu}\bm{\delta}-m^{2}_{\delta}\bm{\delta}^{2}\right). \nonumber
\end{eqnarray}
$U(\sigma)$ is the  potential of $\sigma$ field,
\begin{eqnarray}
U(\sigma)=\left\{
\begin{array}{cc}
\frac{1}{3}g_{2}\sigma^{3}+\frac{1}{4}g_{3}\sigma^{4}& \text{NL} \\
0 & \text{DD}\\
\end{array} \right.
\end{eqnarray}
where NL and DD represent nonlinear and density dependence parameter sets respectively.
 $\mathcal{L}_I$ is
\begin{eqnarray}
\label{lag_i}
\mathcal{L}_I&=&\mathcal{L}_{NN}+\mathcal{L}_{\Delta \Delta}+\mathcal{L}_{N\Delta}\nonumber\\
&=&\Gamma_{\sigma NN}\bar{\Psi}\Psi\sigma-\Gamma_{\omega NN}\bar{\Psi}\gamma_{\mu}\Psi\omega^{\mu}-\Gamma_{\rho NN}\bar{\Psi}\gamma_{\mu}\bm{\tau} \cdot\Psi\bm{\rho}^{\mu}\nonumber\\
&&+\frac{g_{\pi NN}}{m_{\pi}}\bar{\Psi}\gamma_{\mu}\gamma_{5}\bm{\tau} \cdot\Psi\partial^{\mu}\bm{\pi}+\Gamma_{\delta NN}\bar{\Psi}\bm{\tau} \cdot\Psi\bm{\delta}\nonumber\\
&&+\Gamma_{\sigma \Delta \Delta}\bar{\Delta}_{\mu}\Delta^{\mu}\sigma-\Gamma_{\omega \Delta \Delta}\bar{\Delta}_{\mu}\gamma_{\nu}\Delta^{\mu}\omega^{\nu} \nonumber\\
&&-\Gamma_{\rho \Delta\Delta}\bar{\Delta}_{\mu}\gamma_{\nu}\bm{\mbox{T}} \cdot\Delta^{\mu}\bm{\rho}^{\nu}+\frac{g_{\pi \Delta\Delta}}{m_{\pi}}\bar{\Delta}_{\mu}\gamma_{\nu}\gamma_{5}\bm{\mbox{T}} \cdot\Delta^{\mu}\partial^{\nu}\bm{\pi}\nonumber\\
&&+\Gamma_{\delta \Delta\Delta}\bar{\Delta}_{\mu}\bm{\mbox{T}} \cdot\Delta^{\mu}\bm{\delta}+\frac{g_{\pi N\Delta}}{m_{\pi}}\bar{\Delta}_{\mu}\bm{\mathcal{T}}\cdot \Psi\partial^{\mu}\bm{\pi}\nonumber\\
&&+\frac{ig_{\rho N\Delta}}{m_{\rho}}\bar{\Delta}_{\mu}\gamma_{\nu}\gamma_{5}\bm{\mathcal{T}}\cdot \Psi\left(\partial^{\nu}\bm{\rho}^{\mu}-\partial^{\mu}\bm{\rho}^{\nu}\right)+h.c. ~
\end{eqnarray}

 where $\omega_{\mu\nu}$ and $\bm{\rho}_{\mu\nu}$ in Eq.(\ref{lag_f}) are defined by $\partial_{\mu}\omega_{\nu}-\partial_{\nu}\omega_{\mu}$ and $\partial_{\mu}\bm{\rho}_{\nu}-\partial_{\nu}\bm{\rho}_{\mu}$, respectively. Here $\bm{\tau}$ and $\mathbf{T}$ are the isospin matrices of nucleon and $\Delta$ \cite{Machleidt1987,Benmerrouche1989} , and $\bm{\mathcal{T}}$ is the isospin transition matrix between the isospin 1/2 and 3/2 fields \cite{Huber1994} . $\Gamma_{i NN}$  is meson-nucleon  coupling constant
\begin{eqnarray}
\Gamma_{i NN}=\left\{
\begin{array}{cc}
g_{i NN}& \text{NL} \\
g_{i NN}(\rho) & \text{DD}\\
\end{array} \right.
\end{eqnarray}
the values of $\Gamma_{iNN}$ are listed in Table I.

For the coupling constants $\Gamma_{i\Delta\Delta}=\Gamma_{iNN}$, $i=\sigma, \omega, \rho, \delta$,  are the same as Refs.\cite{Song2015,QingfengLi2017,Larionov2003}. The coupling constant $g_{\pi N\Delta}$ is indispensable for describing the $NN\to N\Delta$ cross section, and it is determined by analyzing the $\Delta$-isobar decay width from Ref.~\cite{Dmitriev1986} .

In the uniform rest nuclear matter, the effective momentum can be written as $\textbf{p}_i^*=\textbf{p}_i$ since the spatial components of vector field vanish, i.e., $\mathbf{\Sigma}=0$. Thus, in the mean field approach, the effective energy reads as
\begin{equation}
p_i^{*0}=p^{0}_{i}-\Sigma^{0}_{i},
\end{equation}
and
\begin{eqnarray}
\Sigma^{0}_{i}=\left\{
\begin{array}{cc}
g_{\omega NN}\bar{\omega}^{0}+g_{\rho NN}t_{3,i}\bar{\rho}^{0}_3& \text{NL} \\
\Gamma_{\omega NN}\bar{\omega}^{0}+\Gamma_{\rho NN}t_{3,i}\bar{\rho}^{0}_3+\Sigma^{r} & \text{DD}\\
\end{array} \right.
\end{eqnarray}
where $\Sigma^{r}=\frac{\partial\Gamma_{\omega NN}}{\partial\rho}\bar{\omega}^{0}\rho+\frac{\partial \Gamma_{\rho NN}}{\partial\rho}\bar{\rho}^{0}_3\rho_3-\frac{\partial\Gamma_{\sigma NN}}{\partial\rho}\bar{\sigma}\rho_s-\frac{\partial\Gamma_{\delta NN}}{\partial\rho}\bar{\delta}_3\rho_{s3}$ is the rearrangement term in density dependence RMF sets with:
\begin{eqnarray}
&&\rho_s=\langle  \bar{\Psi}\Psi \rangle= \rho_{s n}+\rho_{s p} \label{rs}\\
&&\rho=\langle  \bar{\Psi}\gamma^0 \Psi \rangle= \rho_{n}+\rho_{p} \label{rv}\\
&&\rho_{s3}=\langle  \bar{\Psi}\tau_3 \Psi \rangle= \rho_{s p}-\rho_{s n} \label{rst}\\
&&\rho_3=\langle  \bar{\Psi}\gamma^0 \tau_3 \Psi \rangle= \rho_{p}-\rho_{n} \label{rvt}.
\end{eqnarray}
Here $t_{3,i}$ is the third component of the isospin of the nucleon and $\Delta$, and i=n, p, $\Delta^{++}$, $\Delta^{+}$, $\Delta^{0}$, $\Delta^{-}$, where $t_{3,n}=-1$, $t_{3,p}=1$, $t_{3,\Delta^{++}}=1$, $t_{3,\Delta^{+}}=\frac{1}{3}$, $t_{3,\Delta^{0}}=-\frac{1}{3}$, $t_{3,\Delta^{-}}=-1$, and $\bar{\rho}^{0}_3=\frac{\Gamma_{\rho NN}}{m^2_\rho}(\rho_{p}-\rho_{n}$).
The Dirac effective masses of nucleon and $\Delta$ read as:
\begin{equation}
m^{*}_{i}=m_{i}+\Sigma^{S}_{i},
\label{eq:efmnd}
\end{equation}
where
\begin{equation}
\Sigma^{S}_{i}=-\Gamma_{\sigma NN}\bar{\sigma}- \Gamma_{\delta NN}t_{3,i}\bar{\delta}_3,
\label{eq:efmnd2}
\end{equation}
and $\bar{\delta}_3=\frac{\Gamma_{\delta NN}}{m^2_{\delta}}(\rho_{sp}-\rho_{sn})$.

\subsection{In-medium $NN\rightarrow N\Delta$ cross section}

Applying quasiparticle approximation \cite{Baym76}, the in-medium cross sections are introduced via the replacement of the vacuum plane waves of the initial and final particles by the plane waves obtained by solution of the nucleon and $\Delta$ equations of motion with scalar and vector fields. In detail, the matrix elements $\mathcal{M}^*$ for the inelastic scattering process $NN\rightarrow N\Delta$ are obtained by replacing the nucleon and $\Delta$ masses and momenta in free space with their effective masses and kinetic momenta \cite{Larionov2003}, i.e., $m \to m^*$ and $p^{\mu}\to p^{* \mu}$.
 As in Ref.~\cite{Larionov2003}, all the calculations performed in this work are for colliding nucleons with their center-of-mass frame coinciding with the nuclear matter rest frame. In this case, the spatial component of the nucleon vector self-energy vanishes.

\begin{figure}[htbp]
\begin{center}
    \includegraphics[scale=0.1]{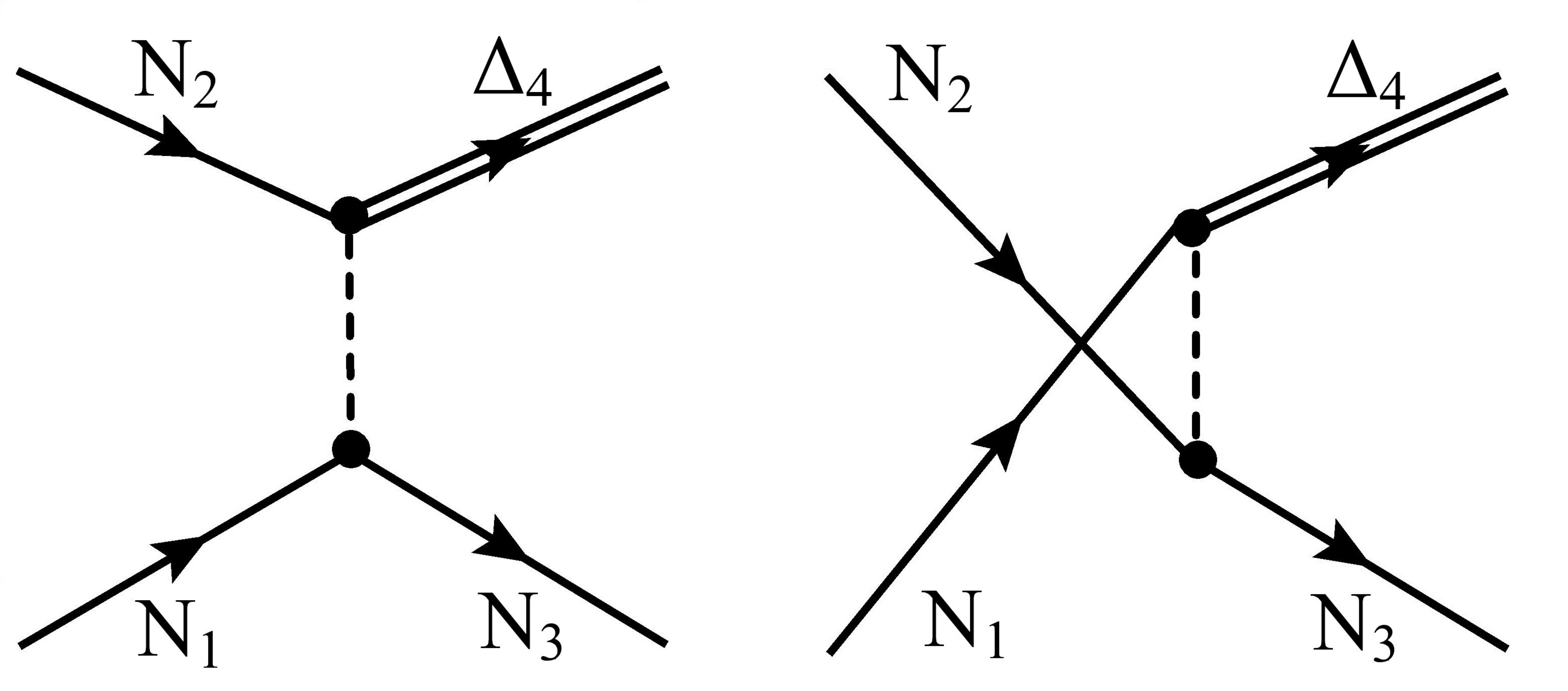}
    \caption{The left diagram is the direct term, and the right one is the exchange term.}\label{fig0}
\end{center}
\end{figure}

The Feynmann diagrams corresponding to the inelastic-scattering $NN\rightarrow N\Delta$ processes are shown in Fig. ~\ref{fig0}, which include the direct and exchange processes. The $\mathcal{M}^*$-matrix for the interaction Lagrangian Eq.~(\ref{lag_i}) can be written by the standard procedure \cite{Huber1994},

\begin{equation}
\mathcal{M}^*=\mathcal{M}_d^{*\pi}-\mathcal{M}_e^{*\pi}+\mathcal{M}_d^{*\rho}-\mathcal{M}_e^{*\rho},
\end{equation}
where
\begin{eqnarray}
\mathcal{M}_d^{*\pi}&=&-i\frac{g_{\pi NN} g_{\pi N\Delta} I_d  }{ m_{\pi}^{2}( Q^{*2}_{d}- m_{\pi}^{2})}[\bar{\Psi}(p_3^* ) \gamma_{\mu}\gamma_5  Q_{d}^{*\mu}  \Psi(p_1^*)]\nonumber\\
&&\times[\bar{\Delta}_{\nu} (p_4^* ) Q_d^{*\nu}  \Psi(p_2^* )]\\
\mathcal{M}_d^{*\rho}&=&i\frac{\Gamma_{\rho NN} g_{\rho N\Delta}I_d}{m_{\rho} }[\bar{\Psi}(p_3^* ) \gamma_{\mu} \Psi(p_1^* )]\\\nonumber
&&\times\frac{g^{\mu\tau}-Q_d^{*\mu} Q_d^{*\tau}/m^{2}_{\rho}}{Q_d^{*2}-m^{2}_{\rho}}\\\nonumber
&&\times[\bar{\Delta}_{\sigma} (p_4^* ) \gamma_{\lambda} \gamma_{5} (Q_d^{*\lambda} \delta_{\sigma\tau}-Q_d^{*\sigma} \delta_{\lambda\tau}) \Psi(p_2^* )]~.
\end{eqnarray}

Here $Q_{d}^{*\mu}=p_{3}^{*\mu}-p_{1}^{*\mu}$ for the direct term, the exchange term $\mathcal{M}^*_e$ is obtained by $p_{1}^{*\mu}\longleftrightarrow p_{2}^{*\mu}$ and $Q_{e}^{*\mu}=p_{3}^{*\mu}-p_{2}^{*\mu}$. The isospin factors $I_d$, $I_e$  can be found in the Ref.~\cite{Huber1994}.

The in-medium $NN\rightarrow N\Delta$ cross section is the in-medium elementary two-body cross section averaged over the mass of $\Delta$ by considering the $\Delta$ as the short-living resonance, and it can be written as:
\begin{eqnarray}
&&\sigma^*_{NN\rightarrow N\Delta}=\int_{m^*_{\Delta,\text{min}}}^{m^*_{\Delta,\text{max}}} dm^*_{\Delta}f(m^*_{\Delta})\tilde{\sigma}^*(m^*_{\Delta}),
\label{eq:xsnd1}
\end{eqnarray}
where $\tilde{\sigma}^*( m^*_{\Delta})$ is the in-medium elementary two-body cross section. In the center-of-mass frame of colliding nucleons, it reads
\begin{eqnarray}
\label{eq:xsnd2}
\tilde{\sigma}^*( m^*_{\Delta})
=\frac{1}{64\pi^2}\int \frac{|\textbf{p}^{*}_{\text{out, c.m.}}|}{\sqrt{s^*_{\text{in}}}\sqrt{s^*_{\text{out}}}|\textbf{p}^{*}_{\text{in, c.m.}}|} \overline{|\mathcal{M}^*|^2}  d\Omega,
\end{eqnarray}
where $\textbf{p}^{*}_{\text{in, c.m.}}$ and $\textbf{p}^{*}_{\text{out, c.m.}}$ are the momenta of incoming (1 and 2) and outgoing particles (3 and 4), and $s^*_{\text{in}}=(p^*_1+p^*_2)^2$, and $s^*_{\text{out}}=(p^*_3+p^*_4)^2$.

Here $\overline{|\mathcal{M}^*|^2}=\frac{1}{(2s_{1}+1)(2s_2+1)}\sum\limits_{s_{1}s_{2}s_{3}s_{4}}|\mathcal{M}^*|^2$ is,
\begin{eqnarray}
&&\sum\limits_{s_{1}s_{2}s_{3}s_{4}}|\mathcal{M}^*|^2 \nonumber\\
&&=\sum\limits_{s_{1}s_{2}s_{3}s_{4}} \{ |\mathcal{M}_d^{*\pi}|^2-\mathcal{M}_d^{*\pi \dagger}\mathcal{M}_e^{*\pi}-\mathcal{M}_e^{*\pi \dagger}\mathcal{M}_d^{*\pi}+|\mathcal{M}_e^{*\pi}|^2\nonumber\\
&& +|\mathcal{M}_d^{*\rho}|^2-\mathcal{M}_d^{*\rho \dagger}\mathcal{M}_e^{*\rho}-\mathcal{M}_e^{*\rho \dagger}\mathcal{M}_d^{*\rho}+|\mathcal{M}_e^{*\rho}|^2 \nonumber\\
&&+\mathcal{M}_d^{*\pi \dagger}\mathcal{M}_d^{*\rho}-\mathcal{M}_d^{*\pi \dagger}\mathcal{M}_e^{*\rho}-\mathcal{M}_e^{*\pi \dagger}\mathcal{M}_d^{*\rho}+\mathcal{M}_e^{*\pi \dagger}\mathcal{M}_e^{*\rho} \nonumber\\
&&+\mathcal{M}_d^{*\rho \dagger}\mathcal{M}_d^{*\pi}-\mathcal{M}_d^{*\rho \dagger}\mathcal{M}_e^{*\pi}-\mathcal{M}_e^{*\rho \dagger}\mathcal{M}_d^{*\pi}+\mathcal{M}_e^{*\rho \dagger}\mathcal{M}_e^{*\pi}\}. \nonumber\\\label{eq:xsnd3}
\end{eqnarray}
where the $\mathcal{M}^*$-matrix is from exchange by $\pi$ and $\rho$ mesons in Ref.~\cite{Cui2018} .
Here, we only show the direct term as an example for $\pi$ mesons, i.e., $\sum\limits_{s_{1}s_{2}s_{3}s_{4}} |\mathcal{M}_d^{*\pi}|^2$:
\begin{eqnarray}
\label{eq:xsnd4}
&&\sum\limits_{s_{1}s_{2}s_{3}s_{4}} |\mathcal{M}_d^{*\pi}|^2 =\left(\frac{g_{\pi NN} g_{\pi N\Delta} I_d  }{ m_{\pi}^{2}( Q^{*2}_{d}- m_{\pi}^{2})}\right)^2\nonumber\\
&&\times\sum\limits_{s_{1}s_{2}s_{3}s_{4}}[\Psi(p_1^*)\bar{\Psi}(p_1^* ) \gamma_{\mu}\gamma_5  Q_{d}^{*\mu}  \Psi(p_3^*)\bar{\Psi}(p_3^* )\gamma_{\sigma}\gamma_5  Q_{d}^{*\sigma}]\nonumber\\
&&\times[\Psi(p_2^* )\bar{\Psi} (p_2^* ) Q_d^{*\nu}  \Delta_{\nu} (p_4^* )\bar{\Delta}_{\tau} (p_4^* ) Q_d^{*\tau}]\nonumber\\
&&=\left(\frac{g_{\pi NN} g_{\pi N\Delta} I_d  }{ m_{\pi}^{2}( t^{*}- m_{\pi}^{2})}\right)^2\nonumber\\
&&\times\frac{2 (m^*_{N_{1}} + m^{*}_{N_{3}})^2((m^{*}_{N_{1}} - m^{*}_{N_{3}})^2 - t^*)}{3 m^{*2}_{\Delta_{4}}}\nonumber\\
&&\times\left((m^{*}_{\Delta_{4}}-m^{*}_{N_{2}})^2 - t^*\right) \left((m^{*}_{N_{2}} +  m^{*}_{\Delta_{4}})^2 - t^*\right)^2
\end{eqnarray}
where $t=Q^{*2}_{d}$, for $|\mathcal{M}_e^{*\pi}|^2$ is $N_{1}\longleftrightarrow N_{2}$. With the energy-momentum conservation, $p^\mu_{1}+p^\mu_{2}= p^\mu_{3}+p^\mu_{4}$ can be expressed as $p^{*\mu}_{1}+\Sigma^{\mu}_{1}+ p^{*\mu}_{2}+\Sigma^{\mu}_{2}= p^{*\mu}_{3}+\Sigma^{\mu}_{3}+p^{*\mu}_{4}+\Sigma^{\mu}_{4}$, $p^{*\mu}_{1}+ p^{*\mu}_{2}= p^{*\mu}_{3}+p^{*\mu}_{4}-\Delta\Sigma^{\mu}$, here $\Delta\Sigma^{\mu}=\Sigma^{\mu}_{1}+\Sigma^{\mu}_{2}-\Sigma^{\mu}_{3}-\Sigma^{\mu}_{4}$ is the kinetic momentum change between the initial and final states, and the effective energy changes are expressed as $\Delta \Sigma^0=\Sigma^{0}_{1}+\Sigma^{0}_{2}-\Sigma^{0}_{3}-\Sigma^{0}_{4}$.
Consequently, $p^{*0}_1+p^{*0}_2$ may differ from $p^{*0}_3+p^{*0}_4$, and $s^*_{in}\ne s^*_{out}$ in Eq.~(\ref{eq:xsnd2}),  and they are related according to the following relationship,
\begin{equation}
\label{sinstar}
\sqrt{s}=\sqrt{s^*_{in}}+\Sigma^0_{N_1}+\Sigma^0_{N_2}=\sqrt{s^*_{out}}+\Sigma^0_{N_3}+\Sigma^0_{\Delta_4}.
\end{equation}
It is derived from
\begin{eqnarray}
\label{eq:sin}
s&=&(p_{N_{1}}+p_{N_{2}})^2\nonumber\\
&=&(\sqrt{m^{*2}_{N_{1}}+\mathbf{p}^{*2}_{N_{1}}}+\sqrt{m^{*2}_{N_{2}}+\mathbf{p}^{*2}_{N_{2}}}+\Sigma^0_{N_1}+\Sigma^0_{N_2})^2-(\mathbf{p}^*_{N_{1}}+\mathbf{p}^*_{N_2})^2 \nonumber\\
&=&(p_{N_{3}}+p_{\Delta_4})^2\nonumber\\
&=&(\sqrt{m^{*2}_{N_{3}}+\mathbf{p}^{*2}_{N_{3}}}+\sqrt{m^{*2}_{\Delta_4}+\mathbf{p}^{*2}_{\Delta_4}}+\Sigma^0_{N_3}+\Sigma^0_{\Delta_4})^2-(\mathbf{p}^*_{N_{3}}+\mathbf{p}^*_{\Delta_4})^2\\\nonumber
\end{eqnarray}
where $\mathbf{p}^*_{N_1}=-\mathbf{p}^*_{N_2}$ and $\mathbf{p}^*_{N_3}=-\mathbf{p}^*_{\Delta_4}$ in the center-of-mass frame.

And, $m^*_{\Delta,\text{min}}$ in the formula of the cross section is determined by the $\Delta \rightarrow N+ \pi$ in isospin asymmetric nuclear matter as in  Refs.~\cite{ZhenZhang2017,Cui2018} when both $N$ and $\pi$ are at rest. The modification of scalar and vector self-energies in this isospin exchange process should also be considered. Thus,
\begin{equation}
m^*_{\Delta,\text{min}}=m^*_{N}+m_\pi-\Delta\Sigma_d^0,
 \end{equation}
 with  $\Delta\Sigma_d^0=\Sigma_\Delta^0-\Sigma_{N}^0$.
The $m^*_{\Delta,\text{max}}$ is evaluated from $NN\to N\Delta$ for producing $N$ and $\Delta$ at rest, and it leads to
\begin{equation}
m^*_{\Delta,\text{max}}=\sqrt{s}-m^*_{N_{3}}-\Sigma^0_{N_{3}}-\Sigma^0_{\Delta_4}.
\end{equation}

The in-medium $\Delta$ mass distribution $f(m^*_\Delta)$ is another important ingredient of in-medium $NN\rightarrow N\Delta$ cross section for which proper energy conservation is also necessary since  $f(m^*_\Delta)$ is related to the $\Delta\rightarrow N+\pi$ process in isospin asymmetric
nuclear matter. In this paper, the spectral function of  $\Delta$ is taken as in Ref.~\cite{Larionov2003} ,
\begin{equation}
\label{eq:bt}
f(m^*_{\Delta})=\frac{2}{\pi}\frac{m^{* 2}_{\Delta}\Gamma(m^{*}_{\Delta})}{(m^{*2}_{0,\Delta}-m^{*2}_{\Delta})^2+m^{*2}_{\Delta}\Gamma^2(m^{*}_{\Delta}) }.
\end{equation}
Here, $m^*_{0,\Delta}$ is the effective pole mass of $\Delta$. The decay width $\Gamma(m^*_\Delta)$ is taken as the parameterization form \cite{Larionov2003}
\begin{eqnarray}
\label{eq:gama}
\Gamma(m^{*}_{\Delta})&=&\Gamma_{0}\frac{q^{3}(m^{* }_{\Delta},m^*_N,m^*_\pi)}{q^{3}(m^{*}_{0,\Delta},m^*_N,m^*_\pi)}\frac{q^{3}(m^{*}_{0,\Delta},m^*_N,m^*_\pi)+\eta^2}{q^{3}(m^{* }_{\Delta},m^*_N,m^*_\pi)+\eta^2}\frac{m^{*}_{0,\Delta}}{m^{*}_{\Delta}},
\end{eqnarray}
where
\begin{eqnarray}
\label{eq:qm123}
&&q(m^{*}_\Delta,m^*_{N},m^*_\pi)=\sqrt{\frac{\left((m^*_\Delta+\Sigma^{0}_{\Delta}-\Sigma^{0}_{N})^2+m_{N}^{*2}-m_{\pi}^{* 2}\right)^2}
{4(m^{*}_\Delta+\Sigma^{0}_{\Delta}-\Sigma^{0}_{N})^2}-m_{N}^{*2}}.
\end{eqnarray}

The coefficients of $\Gamma_0$=0.118 GeV and $\eta$=0.2 GeV/$c$ are used in the above parameterization formula.

\begin{table*}[htbp]
\begin{center}
\caption{The parameters used in the effective Lagrangian, $g_{\pi NN}$=1.008, $g_{\pi N\Delta}$=2.202, $m_{\pi}$=138, $m_{N}$=939, $m_{0,\Delta}$=1232 (all masses are in MeV), $g_{2}/g_{\sigma NN}^3$=0.03302 fm$^{-1}$, $g_{3}/g_{\sigma NN}^4$=-0.00483 (NL$\rho$-$\Delta$ and NL$\rho\delta$-$\Delta$), $\Lambda_{\pi NN}$=1000MeV. The coupling constants $\Gamma_{iNN}$ and $g_{iN\Delta}$ are dimensionless.}\label{table-para}
\resizebox{\textwidth}{30mm}{
\begin{threeparttable}
\begin{tabular}{c   c   c   c  c   c   c}
  \hline
  \hline
    & NL$\rho$-$\Delta$\cite{Liu2005} & DDME2-$\Delta$\cite{Lalazissis2005}\tnote{a} & DDRH$\rho$-$\Delta$\cite{Gaitanos2004}\tnote{a} & NL$\rho\delta$-$\Delta$ \cite{Liu2005}& DDME$\delta$-$\Delta$\cite{Roca2011}\tnote{a} & DDRH$\rho\delta$-$\Delta$\cite{Gaitanos2004}\tnote{a} \\
    \hline
    $m_{\sigma}$ (MeV) & 550 & 566 & 550 & 550 & 566 & 550  \\
    $m_{\omega}$ (MeV)& 783 &  783 & 783 & 783 &  783 & 783 \\
    $m_{\rho}$ (MeV)& 770 & 769 & 763 & 770 & 769 & 763 \\
    $m_{\delta}$ (MeV) & 980 &983 & 980 & 980 &983 & 980\\
   $\Gamma_{\sigma NN}$ &8.9679 & 10.5396 & 10.7286  &8.9679 & 10.3313 & 10.7286 \\
   $\Gamma_{\omega NN}$  &9.2408 &  13.0189 & 13.2902  &9.2408 &  12.2905 & 13.2902 \\
   $\Gamma_{\rho NN}$ &3.8033 &  3.6836 & 3.5655 &6.9256 &  6.3119 & 5.8284 \\
   $\Gamma_{\delta NN}$ &0 &  0 &  0 &7.8525 &  7.1515 &  7.6009\\
    $\Lambda_{\pi N\Delta}$ (MeV) &419 &  410 & 410 &410 &  416 & 417 \\
   $\Lambda_{\rho NN}$ (MeV) &1000 &  1000 & 1000 &1000 &  650 & 580 \\
   $m^{*}_{N}/m_{N}$ &0.75 &  0.57 & 0.55 &0.75 &  0.61 & 0.55 \\
 $m^{*}_{\Delta}/m_{\Delta}$ & 0.809 & 0.674  & 0.661& 0.809 & 0.702  & 0.661 \\
   $\Delta m^*_N$ \tnote{b}&0&0&0 &0.0312 &  0.0236 & 0.0265 \\
   $\Delta m^*_\Delta$\tnote{b}&0&0&0 & 0.0079  & 0.0060  & 0.0068  \\
   \hline
   \hline	 		 		 	 	 			 	 	
\end{tabular}
\begin{tablenotes}\small{
\item[a] The density dependent coupling constants of DDME2-$\Delta$,  DDRH$\rho$-$\Delta$,  DDME$\delta$-$\Delta$ and DDRH$\rho\delta$-$\Delta$  are for $\rho=\rho_0$.
\item[b] Here $\Delta m^*_N=\frac{m^{*}_{p}-m^{*}_{n}}{m_{N}}$ and $\Delta m^*_\Delta=\frac{m^{*}_{\Delta^{++}}-m^{*}_{\Delta^+}}{m_{\Delta}}$ at isospin asymmetry  $\alpha$=0.2.}
\end{tablenotes}
\end{threeparttable}}
\end{center}
\end{table*}

\begin{table*}[htbp]
\begin{center}
\caption{The parameters for  DDME2-$\Delta$, DDME$\delta$-$\Delta$, DDRH$\rho$-$\Delta$ and DDRH$\rho\delta$-$\Delta$ are given by $g_i(\rho)=g_i(\rho_0)f_i(x)$, for i=$\sigma, \omega, \rho, \delta$, where $x=\rho/\rho_0$.}\label{table-para2}
\begin{threeparttable}
\begin{tabular}{c  c   c   c   c  c   c   c   c}
  \hline
  \hline
    &   \multicolumn{4}{c|}{DDME2-$\Delta$\cite{Lalazissis2005}\tnote{a}} & \multicolumn{4}{c}{DDME$\delta$-$\Delta$\cite{Roca2011}\tnote{b}}  \\
    \hline
    meson & $\sigma$ & $\omega$ & \multicolumn{2}{c|}{$\rho$} & $\sigma$ & $\omega$ & $\rho$ & $\delta$ \\
     $g_{i}(\rho_0)$ & 10.5396 &  13.0189 & \multicolumn{2}{c|}{3.6836 } & 10.3325 &  12.2904 & 7.1520 & 6.3128\\
    $a_{i}$ & 1.3881 &  1.3892 & \multicolumn{2}{c|}{0.5647 } & 1.3927 &  1.4089 & 1.5178 & 1.8877 \\
    $b_{i}$ & 1.0943 & 0.9240 & \multicolumn{2}{c|}{ } & 0.1901 & 0.1698 & 0.3262 & 0.0651 \\
    $c_{i}$ & 1.7057 & 1.4620 & \multicolumn{2}{c|}{ } & 0.3679 & 0.3429 & 0.6041& 0.3469 \\
    $d_{i}$ & 0.4421 &0.4775 & \multicolumn{2}{c|}{ } & 0.9519 & 0.9860 & 0.4257 & 0.9417 \\
    $e_{i}$ & 0.4421 &0.4775 & \multicolumn{2}{c|}{} & 0.9519 & 0.9860 & 0.5885 & 0.9737\\
   \hline
   &   \multicolumn{4}{c|}{DDRH$\rho$-$\Delta$\cite{Gaitanos2004}\tnote{c}} & \multicolumn{4}{c}{DDRH$\rho\delta$-$\Delta$\cite{Gaitanos2004}\tnote{c}}  \\
    \hline
    meson & $\sigma$ & $\omega$ & \multicolumn{2}{c|}{$\rho$} & $\sigma$ & $\omega$ & $\rho$ & $\delta$ \\
     $g_{i}(\rho_0)$ & 10.7285 &  13.2902 & \multicolumn{2}{c|}{3.5870 } & 10.7285 &  13.2902 & 5.8635 & 7.5896\\
    $a_{i}$ & 1.3655 & 1.4025 & \multicolumn{2}{c|}{0.0953 } & 1.3655 &  1.4025 & 0.0953 & 0.0198 \\
    $b_{i}$ & 0.2261 & 0.1726 & \multicolumn{2}{c|}{2.1710 } & 0.2261 & 0.1726 & 2.1710 & 3.4732 \\
    $c_{i}$ & 0.4097 & 0.3443 & \multicolumn{2}{c|}{ 0.0534} & 0.4097 & 0.3443  & 0.0534& -0.0908 \\
    $d_{i}$ & 0.9020 &0.9840 & \multicolumn{2}{c|}{ 17.8431} & 0.9020 &0.9840  & 17.8431 & -9.8110 \\
    \hline
    \hline		 		 	 	 			 	 	
\end{tabular}
\begin{tablenotes}\small{
\item[a] The density dependent coupling constants for DDME2-$\Delta$ are proposed as $f_i=a_i \frac{1+b_i(x+d_i)^2}{1+c_i(x+e_i)^2}$ for $i=\sigma, \omega$, and $f_i=exp[-a_i(x-1)]$ for $i=\rho$.
\item[b] The density dependent coupling constants for DDME$\delta$-$\Delta$ are reads $f_i=a_i \frac{1+b_i(x+d_i)^2}{1+c_i(x+e_i)^2}$ for $i=\sigma, \omega,\rho,\delta$.
\item[c] The density dependent coupling constants for DDRH$\rho$-$\Delta$ and DDRH$\rho\delta$-$\Delta$ are read as $f_i=a_i \frac{1+b_i(x+d_i)^2}{1+c_i(x+d_i)^2}$ for $i=\sigma, \omega$, and $f_i=a_iexp[-b_i(x-1)]-c_i(x-d_i)$ for $i=\rho,\delta$.}
\end{tablenotes}
\end{threeparttable}
\end{center}
\end{table*}

The form factors are adopted to effectively consider the contributions from high-order terms and the finite size of baryons \cite{Huber1994,Vetter1991} , which read
\begin{eqnarray}
\label{factorn}
&&F_N (t^*)=\frac{\Lambda_N^2}{\Lambda_N^2-t^*}  exp\left(-b\sqrt{s^*-4m_N^{* 2}}\right)\\
&&F_{\Delta}(t^*)=\frac{\Lambda_{\Delta}^2}{\Lambda_{\Delta}^2-t^*}.
\end{eqnarray}
Here $F_N (t^* )$ is the form factor for nucleon-meson-nucleon , and $F_\Delta (t^*)$  for  nucleon-meson-$\Delta$ coupling, $b$=0.046 GeV$^{-1}$ for both $\rho NN$ and $\pi NN$ coupling.
The cutoff parameter $\Lambda_{\pi N N}\approx 1$ GeV, $\Lambda_{\rho N N}$ and $\Lambda_{\pi N \Delta}$ are determined by best fitting the data of $NN\rightarrow N\Delta$ cross section in free space \cite{Baldini1987} ranging from $\sqrt{s}$=2.0 to 5 GeV. In the Table.~\ref{table-para}, $\Lambda_{\rho N \Delta}$ is determined based on the relationship $\Lambda_{\rho N \Delta}=\Lambda_{\rho NN}\frac{\Lambda_{\pi N\Delta}}{\Lambda_{\pi NN}}$ as in \cite{Huber1994} . Concerning the coupling constant $g_{\rho N\Delta}$, we use $g_{\rho N\Delta}\approx\frac{\sqrt{3}}{2} \Gamma_{\rho NN} \frac{m_{\rho}}{m_N}$ which are derived from the static quark model \cite{Huber1994}. The corresponding nuclear matter parameters at normal density are listed in the lower part of Table~\ref{table-para}. Here the effective mass splittings $\Delta m^*_{N}$ and $\Delta m^*_{\Delta}$, are taken at isospin asymmetry $\alpha=(\rho_n-\rho_p)/\rho =0.2$, and the effective masses and mass splitting as function of density can be found in Refs.\cite{Cui2019,Gaitanos2004}. And the Table.~\ref{table-para2} lists the parameters of the coupling constants for DDME2-$\Delta$\cite{Lalazissis2005}, DDME$\delta$-$\Delta$\cite{Roca2011}, DDRH$\rho$-$\Delta$\cite{Gaitanos2004} and DDRH$\rho\delta$-$\Delta$\cite{Gaitanos2004}.

\section{Results and discussions}
\label{xs}

In order to investigate the $\delta$ meson influences  on the splitting of the isospin dependence of the in-medium $NN\rightarrow N\Delta$ cross sections,  the selected parameter sets are adjusted to reproduce the experimental data  \cite{Baldini1987} of $pp\rightarrow n\Delta^{++}$ cross sections, which are listed in Table.\ref{table-para}.

The in-medium $NN\rightarrow N\Delta$ cross sections in nuclear matter are estimated with $m \to m^*$ and $p^{\mu}\to p^{*\mu}$, we can see that the  Dirac effective masses and  momenta  of nucleons and $\Delta$s directly influence the results of  $\sigma^*_{NN\rightarrow N\Delta }$ based on Eq~.\ref{eq:xsnd1}-\ref{eq:xsnd4}.

\begin{figure}[htbp]
\begin{center}
    \includegraphics[scale=0.37]{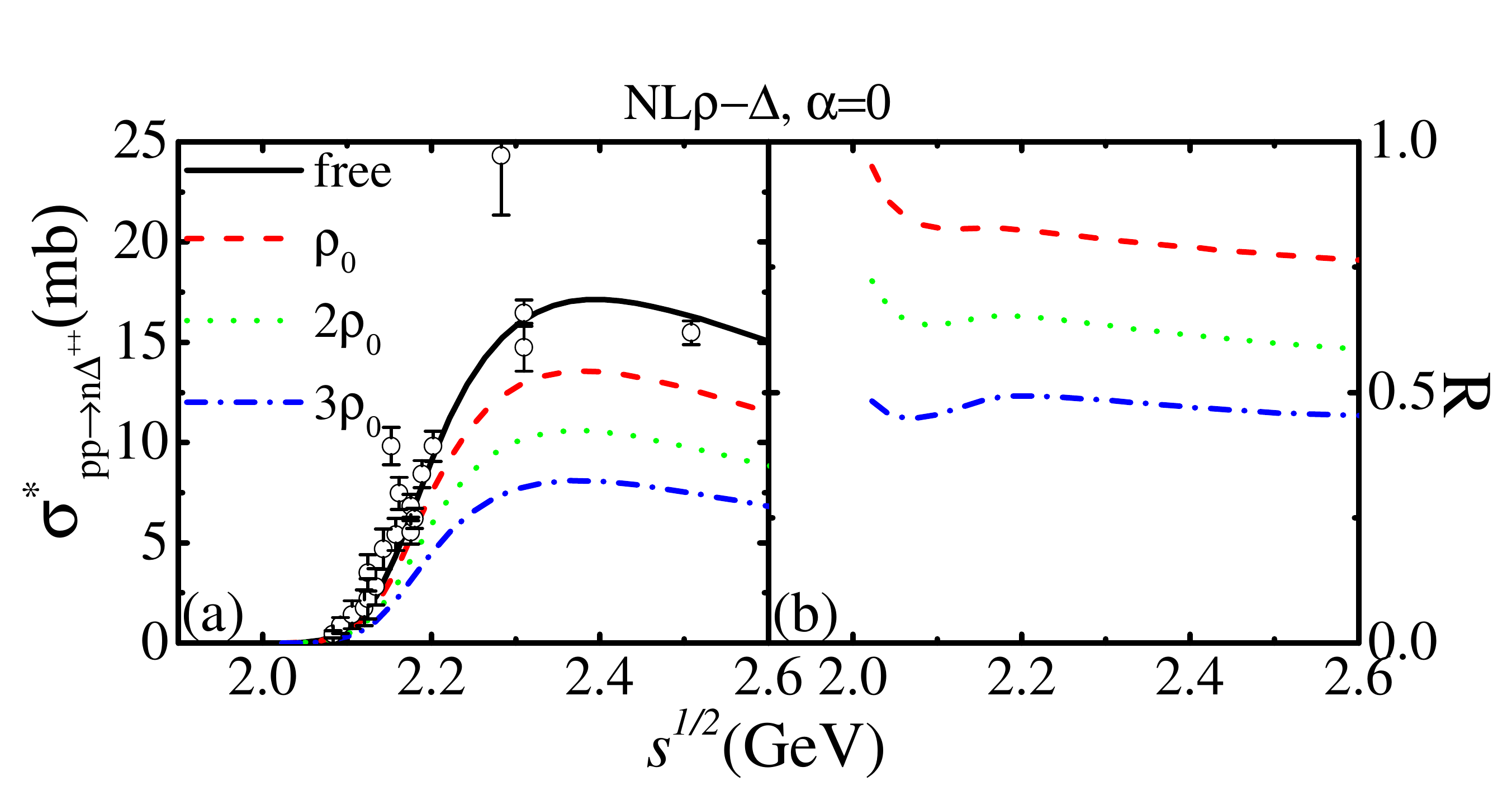}
    \caption{(a) $\sigma^*_{NN\rightarrow N\Delta}$ as a function of $\sqrt{s}$ and the experimental data are taken from Ref.~\cite{Baldini1987}; (b) $R$ as a function of $\sqrt{s}$ in symmetric nuclear matter for $\rho=0$, $\rho_0$ and $2\rho_0$.}\label{fig3}
\end{center}
\end{figure}

 Fig.~\ref{fig3} shows the values of in-medium $pp\rightarrow n\Delta^{++}$ cross section  (panel(a)) and medium correction factor $R=\sigma^*_{NN\rightarrow N\Delta}/\sigma_{NN\rightarrow N\Delta}$  ( panel (b)) for $NL\rho$-$\Delta$ in symmetric nuclear matter, where other parameter sets give the similar results.  Compared to  $pp\rightarrow n\Delta^{++}$ cross section in free space, the in-medium $pp\rightarrow n\Delta^{++}$ cross sections in symmetric nuclear matter decrease with the density increasing. It is mainly due to the  fact that elementary two-body cross section $\sigma^*(m_\Delta^*)$ decreases with the reduction of $m_N^*$ and $m_{0,\Delta}^*$ as found in Ref.~\cite{QingfengLi2017} . Since there is no isospin splitting of effective masses in symmetric nuclear matter,
the in-medium cross section for $nn\rightarrow p\Delta^{-}$ is equal to $pp\rightarrow n\Delta^{++}$, and other channels can be obtained by the production of the isospin Clebsch-Gordan coefficients, which is 1/3 of $\sigma^*_{pp\rightarrow n\Delta^{++}}$. Thus, the values of $R$ are exactly the same for  all the channels of $NN\rightarrow N\Delta$ in symmetric nuclear matter.

As one can see from panel (b) of Fig.~\ref{fig3}, the  medium correction factors  are more sensitive to the total energy of the collision particles $\sqrt{s}\leq2.1$ GeV or $E_{\mathrm{beam}}\leq$0.4 GeV at the same nuclear density, so studying the $R$ of $NN\rightarrow N\Delta$ in-medium cross section  at the lower energy is more important. In this work, we will focus on the in-medium cross section near the threshold energy, i.e., $E_{\mathrm{beam}}=$0.4 GeV.

\begin{figure}[htp]
\begin{center}
\includegraphics[scale=0.38]{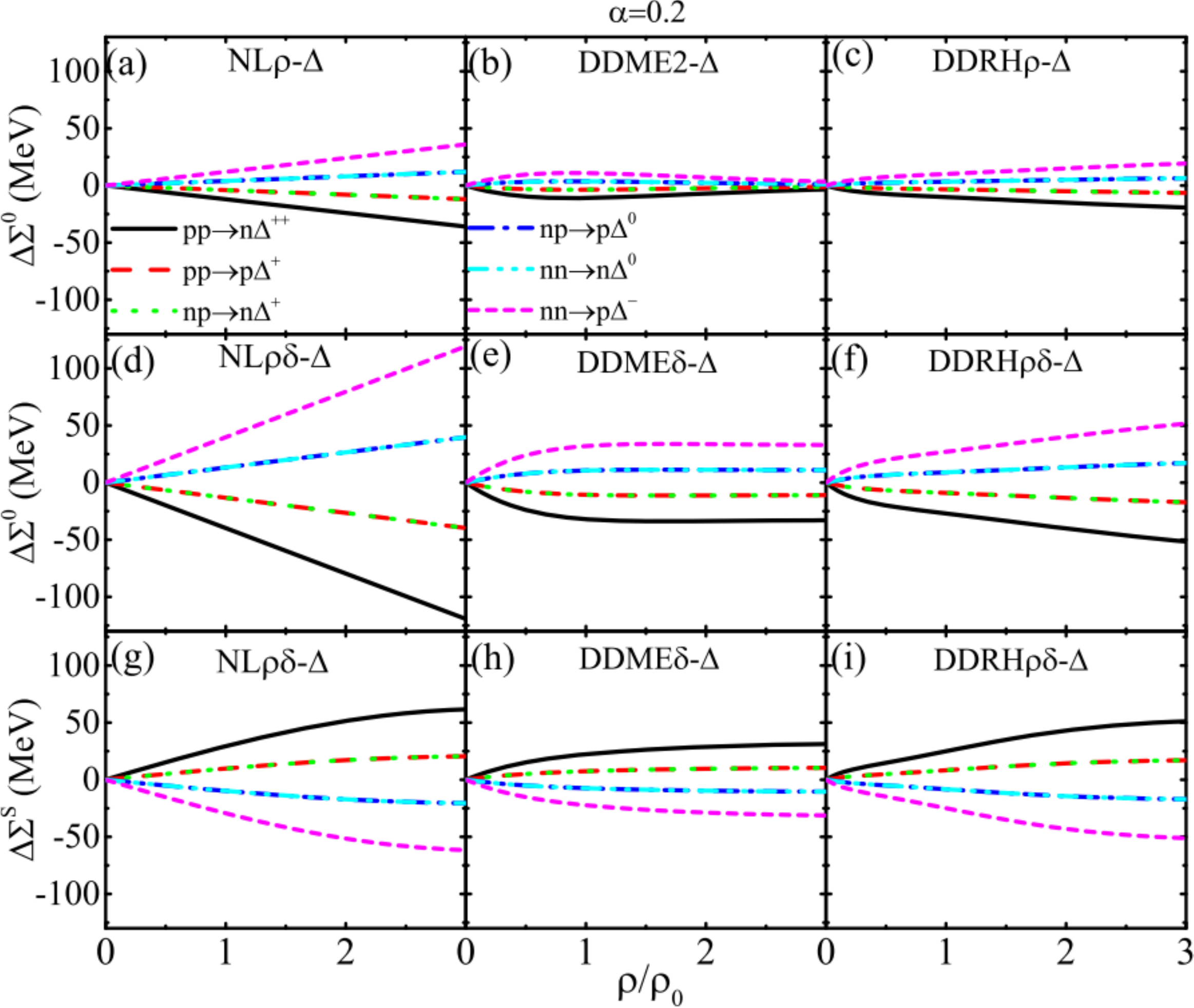}
\caption{The upper and  middle  panels are the vector self-energies changes $\Delta\Sigma^{0}$   as the function of density for different channels of $NN\rightarrow N\Delta$ in without-$\delta$ sets(NL$\rho$-$\Delta$, DDME2-$\Delta$, DDHR$\rho$-$\Delta$) and with-$\delta$ sets(NL$\rho\delta$-$\Delta$, DDME$\delta$-$\Delta$, and DDHR$\rho\delta$-$\Delta$), and bottom panel corresponds to the scalar self-energies changes $\Delta\Sigma^{S}$ in with-$\delta$ sets in asymmetric nuclear matter with  $\alpha=0.2$. }\label{fig5}
\end{center}
\end{figure}

As presented in Eq~.\ref{eq:xsnd1}-\ref{eq:sin}, the cross sections depend on the effective mass and effective energy of nucleons and $\Delta$s, and the isospin splitting of effective mass and effective energy cause the splitting of in-medium correction factor\cite{Cui2018} . Thus, we first investigate the vector self-energies changes $\Delta\Sigma^{0}$ as function of density for different channels of $NN\rightarrow N\Delta$ by using the parameter sets with or without $\delta$ meson, i.e.,  NL$\rho$-$\Delta$, DDME2-$\Delta$,  DDHR$\rho$-$\Delta$, NL$\rho\delta$-$\Delta$, DDME$\delta$-$\Delta$, and DDHR$\rho\delta$-$\Delta$ in isospin asymmetric nuclear matter with $\alpha=0.2$. The results are presented   in the panel (a)-(f) of Fig.~\ref{fig5}. And the scalar self-energies changes $\Delta\Sigma^{S}$ as the function of density for different channels of $NN\rightarrow N\Delta$ with  NL$\rho\delta$-$\Delta$, DDME$\delta$-$\Delta$, and DDHR$\rho\delta$-$\Delta$  are described in panel (g), (h) and (i). In panels (a)-(c) of Fig.~\ref{fig5}, we can see that the splitting  of $\Delta\Sigma^{0}$ for different channels in NL$\rho$-$\Delta$  is relatively large compared to  DDME2-$\Delta$ and DDHR$\rho$-$\Delta$ sets from the results of the without-$\delta$ meson sets. For the with-$\delta$ sets, the splitting amplitude of both $\Delta\Sigma^{0}$ and $\Delta\Sigma^{S}$ are larger in NL$\rho\delta$-$\Delta$ compared to DDME$\delta$-$\Delta$ and DDHR$\rho\delta$-$\Delta$ sets as shown in panels (d)-(i) of Fig.~\ref{fig5}. In addition, we observe that the values of  $\Delta\Sigma^{0}$ and $\Delta\Sigma^{S}$ have the opposite sign for different channels of $NN\rightarrow N\Delta$.

\begin{figure}[htbp]
\begin{center}
    \includegraphics[scale=0.35]{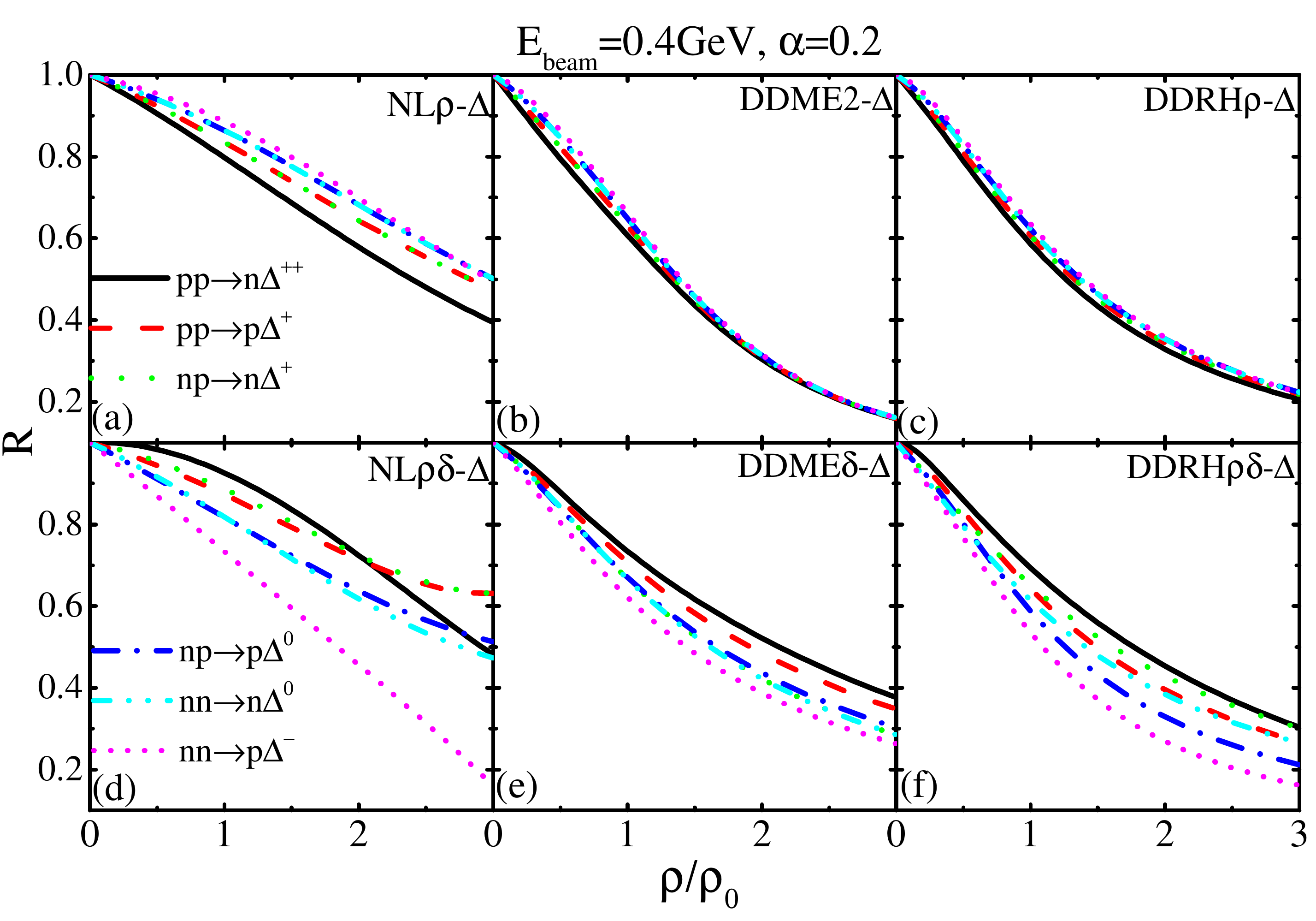}
    \caption{The medium correction factor R at $E_b$=0.4 GeV for NL$\rho$-$\Delta$ , DDME2-$\Delta$ , and DDHR$\rho$-$\Delta$, NL$\rho\delta$-$\Delta$, DDME$\delta$-$\Delta$, and DDHR$\rho\delta$-$\Delta$  in asymmetric nuclear matter in panel (a),(b),(c),(d),(e), and (f). Here the upper panels are the without-$\delta$ sets, and the bottom panels are the $\delta$ sets, respectively. }\label{fig6}
\end{center}
\end{figure}

 Fig.~\ref{fig6} shows results of the medium correction factor $R$ at $E_{\mathrm{beam}}$=0.4 GeV for different parameter sets in asymmetric nuclear matter, where the solid lines are those by using  the without-$\delta$ sets (NL$\rho$-$\Delta$, DDME2-$\Delta$, DDHR$\rho$-$\Delta$) while the dash lines are those of  with-$\delta$ sets(NL$\rho\delta$-$\Delta$, DDME$\delta$-$\Delta$, and DDHR$\rho\delta$-$\Delta$). In previous work\cite{Mao1994c,Mao1994l,Larionov2001,Larionov2003,QingfengLi2000,QingfengLi2017}, the in-medium $NN\rightarrow N\Delta$ cross sections are split in isospin asymmetric nuclear matter because of the effective mass splitting for nucleons and $\Delta$s by considering the $\delta$ meson. From previous study in Ref.\cite{QingfengLi2017}, the effective masses and in-medium  correction factor $R$ are not split in the case of without $\delta$ meson. However, we find there are still splitting  in-medium correction factors  for different channels without considering $\delta$ meson in  asymmetric nuclear matter plotted in Fig.~\ref{fig6}. Especially, the splitting effect is more clear for NL$\rho$-$\Delta$ set, it is also can be seen from Fig.~\ref{fig6} that  $R_{pp\rightarrow n\Delta^{++}}<R_{Np\rightarrow N\Delta^{+}}<R_{Nn\rightarrow N\Delta^{0}}<R_{nn\rightarrow p\Delta^{-}}$, here $N=n$ or $p$. The results can be understood from the  nonvanishing  vector self-energies changes $\Delta\Sigma^0$ due to the existence of $\rho$ meson in asymmetric medium, where  $\Delta\Sigma^0_{pp\rightarrow n\Delta^{++}}>\Delta\Sigma^0_{Np\rightarrow N\Delta^{+}}>\Delta\Sigma^0_{Nn\rightarrow N\Delta^{0}}>\Delta\Sigma^0_{nn\rightarrow p\Delta^{-}}$ in the panels (a)-(c) of Fig.~\ref{fig5}.

The $\Delta \Sigma^0$ comes from the definition of kinetic energy above the threshold energy, i.e. $Q=\sqrt{s}-\sqrt{s_{th}}$, where $\sqrt{s_{th}}=2m_N+m_\pi$. Requiring $Q^*=Q$ ($Q^*=\sqrt{s_{in}^*}-\sqrt{s_{th}^*}$)in  $\Delta$-production process, then :
\begin{eqnarray}
\label{eq:qeff}
Q^*=Q&=&\sqrt{s_{in}}-\sqrt{s_{th}}\nonumber\\
&=&E^*_{N_1}+E^*_{N_2}+\Sigma^0_{N_1}+\Sigma^0_{N_2}\nonumber\\
&&-m^*_{N_3}-m^*_{\Delta,min}-\Sigma^0_{N_3}-\Sigma^0_{\Delta}\nonumber\\\nonumber
&\simeq &(E^*_{N_1}-m^*_{N_1})+(E^*_{N_2}-m^*_{N_2})\nonumber\\
&&+m_{N_1}+m_{N_2}-m_{N_3}-m_{\Delta,min}\nonumber\\
&&+\Delta\Sigma^S+\Delta\Sigma^0
\end{eqnarray}
with $\sqrt{s^*_{th}}=m^*_{N_3}+m^*_{\Delta,min}+\Sigma^0_{N_3}+\Sigma^0_\Delta$, where $\Delta\Sigma^S=\Sigma^{S}_{1}+\Sigma^{S}_{2}-\Sigma^{S}_{3}-\Sigma^{S}_{\Delta} $, $\Delta\Sigma^0=\Sigma^{0}_{1}+\Sigma^{0}_{2}-\Sigma^{0}_{3}-\Sigma^{0}_{\Delta} $.   For example, the $\Delta\Sigma^0(pp\rightarrow n\Delta^{++})$ is smaller than $\Delta\Sigma^0(nn\rightarrow p\Delta^{-})$, so larger ($E^*-m^*$) is needed for $\sigma^*_{pp\rightarrow n\Delta^{++}}$ than $\sigma^*_{nn\rightarrow p\Delta^{-}}$ at given $\sqrt{s}$ (or $Q$) with $\Delta\Sigma^{S}=0$ by considering the energy conservation of canonical momenta in Ref.~\cite{Cui2018} . This effect  will result in the more reduction of in-medium $pp\rightarrow n\Delta^{++}$ cross section than in $nn\rightarrow p\Delta^{-}$.

In Fig.~\ref{fig6}, one can find the most important  influence on the splitting  of isospin dependent $R$ for different channels of $NN\to N\Delta$ from the difference between the without-$\delta$ sets and the with-$\delta$ sets. In addition to $\Delta\Sigma^{0}$ splitting, the separation of the scalar self-energies changes $\Delta\Sigma^{S}$  for different channels of $NN\rightarrow N\Delta$ and the splitting of  effective masses of nucleon and $\Delta$s should also be considered  in  calculating the in-medium cross section by using the with-$\delta$ parameter sets. By comparing the lines in the middle and bottom panels of Fig.~\ref{fig5}, one can draw the conclusion that $\Delta\Sigma^{0}$ and $\Delta\Sigma^{S}$ make the opposite contributions for the splitting among  the different channels of in-medium $NN\rightarrow N\Delta$ cross sections.  However, both $\Delta\Sigma^{0}$ and $\Delta\Sigma^{S}$ can mainly provide the energy shift contribution by considering the energy conservation of canonical momenta  in asymmetric nuclear matter as Eq.~\ref{eq:qeff}. Meanwhile, for the with-$\delta$ sets,  the dominant influence on $R$ split arises from the effective masses splitting, i.e., $m_p^*>m_n^*$, $m_{0,\Delta^{++}}^*>m_{0,\Delta^{+}}^*>m_{0,\Delta^{0}}^*>m_{0,\Delta^{-}}^*$ in the asymmetric nuclear medium, because the in-medium cross sections of $NN\to N\Delta$ are directly calculated by Eqs.~\ref{eq:xsnd2}-\ref{eq:xsnd4} via the values of $m_N^*$, $m^*_\Delta$ not just the total energy shift effect from vector and scalar self-energies changes.  Thus,  we can see that the splitting among different channels are also more obvious in with-$\delta$ sets than that in without-$\delta$ sets from Fig.~\ref{fig6}. In Fig.~\ref{fig6}, one can also observe that the results follow: $R_{pp \to n\Delta ^{++}} <R_{nn \to p\Delta ^{-}}$ and $R_{NN \to N\Delta ^{+}} <R_{NN \to N\Delta ^{0}}$ in without-$\delta$ sets  which are totally reverse for the results obtained by  using the with-$\delta$ sets in present model calculations, where $R_{pp \to n\Delta ^{++}} >R_{nn \to p\Delta ^{-}}$ and $R_{NN \to N\Delta ^{+}} >R_{NN \to N\Delta ^{0}}$. And the reversal effect of the in-medium correction $R_{NN \to N\Delta }$ are more obvious in NL$\rho$-$\Delta$ and NL$\rho\delta$-$\Delta$ than other parameter sets in this work. The order of  $R_{NN\to N\Delta}$  may change the ratio of charged $\pi$ yields $Y(\pi^-)/Y(\pi^+)$($\pi^-/\pi^+$) in HIC since the $\pi$ is directly obtained from the decay of $\Delta$s.  Therefore, if one symmetry energy form could be obtained from without-$\delta$ sets, the in-medium correction factor $R$ should be $R_{pp \to n\Delta ^{++}} <R_{nn \to p\Delta ^{-}}$.

\section{Summary}
\label{summary}
In summary, we have studied the influence of $\delta$-meson on  the  isospin splitting of  the in-medium $NN\rightarrow N\Delta$ cross sections in asymmetric nuclear matter based on the one-boson exchange model. Our calculations show that $\sigma^*_{NN\rightarrow N\Delta}$ decreases with the density increasing for all the channels.  In addition, the $\delta$ meson in RMF sets can influence on the    splitting  of $R$ for the in-medium $NN\to N\Delta$ cross sections.  The results show that the correction factors of the  negative isospin $\Delta$ produced $NN\to N\Delta$ in-medium cross sections  are higher than the positive ones, i.e., $R_{pp \to n\Delta ^{++}} < R_{nn \to p\Delta ^{-}}$ and $R_{NN \to N\Delta ^{+}} <R_{NN \to N\Delta ^{0}}$ by using the without-$\delta$ sets.  By including the $\delta$ meson, it appears the  totally opposite results in the $R$ for different channels, i.e., $R_{pp \to n\Delta ^{++}} > R_{nn \to p\Delta ^{-}}$ and $R_{pp \to n\Delta ^{++}} > R_{nn \to p\Delta ^{-}}$. The dramatic  different results  between two type sets obtained in this work need further study as the coupling constants for $\Delta$-meson-$\Delta$ are largely unclear.  The present results provide a theoretical suggestion on how to modify the isospin dependent in-medium correction factor R which should be correlated to the mean field potential, and will be helpful
for reducing the theoretical uncertainties on the prediction of pion observables.

\section*{Acknowledgements}
This work has been supported by National Key R\&D Program of China under Grant No. 2018 YFA0404404, and National Natural Science Foundation of China under Grants No. 11875321, No. 11875323, No. 11875125, No. 11475262, No. 11961141003, No. 11790323, 11790324, No. 11790325, and the Continuous Basic Scientific Research Project (No. WDJC-2019-13, No BJ20002501).

\end{document}